%% file: 0_mainpaper.tex
\documentclass{article}

\usepackage{cite}
\usepackage{bbm}
\usepackage{caption}
\usepackage{siunitx}

\usepackage{microtype}
\usepackage{subfigure}
\usepackage{booktabs} 

\usepackage[accepted]{icml2024} 

\usepackage{amsmath}
\usepackage{amssymb}
\usepackage{mathtools}
\usepackage{amsthm}
\usepackage{caption}
\usepackage{subcaption}
\usepackage{multirow}
\usepackage[final]{graphicx}
\usepackage[capitalize,noabbrev]{cleveref}

\theoremstyle{plain}
\newtheorem{theorem}{Theorem}[section]

\newtheorem{lemma}[theorem]{Lemma}

\theoremstyle{definition}

\theoremstyle{remark}

\usepackage[textsize=tiny]{todonotes}

\icmltitlerunning{Secure Vertical Federated Learning Under Unreliable Connectivity}

\begin{document}

\twocolumn[
\icmltitle{Secure Vertical Federated Learning Under Unreliable Connectivity}



\icmlsetsymbol{equal}{*}

\begin{icmlauthorlist}
\icmlauthor{Xinchi Qiu}{equal,cam}
\icmlauthor{Heng Pan}{equal,cam}
\icmlauthor{Wanru Zhao}{cam}
\icmlauthor{Yan Gao}{cam}
\icmlauthor{Pedro P. B. Gusmao}{sur}
\icmlauthor{William F. Shen}{cam}
\icmlauthor{Chenyang Ma}{ox}
\icmlauthor{Nicholas D.\ Lane}{cam}

\end{icmlauthorlist}

\icmlaffiliation{cam}{Department of Computer Science and Technology, University of Cambridge}
\icmlaffiliation{ox}{Department of Computer Science, University of Oxford}
\icmlaffiliation{sur}{Department of Computer Science, University of Surrey}

\icmlcorrespondingauthor{Xinchi Qiu}{xq227@cam.ac.uk}
\icmlcorrespondingauthor{Heng Pan}{ac.panh99@gmail.com}

\icmlkeywords{Machine Learning, Federated Learning, ICML}

\vskip 0.3in
]



\printAffiliationsAndNotice{\icmlEqualContribution} 

\begin{abstract}
Most work in privacy-preserving federated learning (FL) has focused on horizontally partitioned datasets where clients hold the same features and train complete client-level models independently. However, individual data points are often scattered across different institutions, known as clients, in vertical FL (VFL) settings. Addressing this category of FL necessitates the exchange of intermediate outputs and gradients among participants, resulting in potential privacy leakage risks and slow convergence rates. Additionally, in many real-world scenarios, VFL training also faces the acute issue of client stragglers and drop-outs, a serious challenge that can significantly hinder the training process but has been largely overlooked in existing studies. In this work, we present \emph{vFedSec}, a \textit{first dropout-tolerant} VFL protocol, which can support the most generalized vertical framework. It achieves secure and efficient model training by using an innovative \emph{Secure Layer} alongside an embedding-padding technique. We provide theoretical proof that our design attains enhanced security while maintaining training performance. Empirical results from extensive experiments also demonstrate \emph{vFedSec} is robust to client dropout and provides secure training with negligible computation and communication overhead. Compared to widely adopted homomorphic encryption (HE) methods, our approach achieves a remarkable $\geq 690\times$ speedup and reduces communication costs significantly by $\geq 9.6\times$. 
\end{abstract}

\input{sections/1_intro}

\input{sections/2_related}

\input{sections/3_method}
\input{sections/4_exp}

\input{sections/5_conclusion}



\section*{Impact Statement}
This paper presents work whose goal is to provide a privacy-preserving vertical federated learning framework. There are many potential societal consequences of our work, none of which we feel must be specifically highlighted here.


\bibliography{example_paper}
\bibliographystyle{icml2024}

\newpage
\appendix
\onecolumn
\input{sections/99_appendix}

\end{document}

%% file: sections/1_intro.tex
\section{Introduction}\label{sec:intro}

Federated learning (FL) is a machine learning paradigm that enables individual institutions or devices to train a global model collaboratively without sharing raw and sensitive data 
~\cite{fedavg}. FL can be categorized according to how datasets are partitioned across clients. Most existing works focus on horizontal FL (HFL), where all participants use the same feature space whilst holding different data samples~\cite{yang2019federated}. This partitioning scheme can often be found in \emph{cross-device} setups where clients are often thousands of intermittent mobile or IoT devices holding different data points, each point covering the entire input space~\cite{yu2021toward, qiu2021zerofl}. Another case is vertical FL (VFL) \cite{wei2022vertical}, where data points are partitioned across clients, which means different clients hold different features of the same sample. VFL was historically only considered for the \emph{cross-silo} setups involving a limited number of participating clients with generally reliable connections. In such setups, participating clients could be institutions holding complementary information on the same data points. An illustrative example is different hospitals each hold different clinical data of the same patient.

However, assuming all participating clients to have a constantly reliable connection with the server dangerously oversimplifies the challenges faced by VFL in real-world scenarios, either due to the specific setup of the system or the practical difficulties of maintaining reliable connectivity due to physical or geographical restrictions. The former could be found in systems where models are jointly trained by data from institutional clients, which have generally reliable connections with the server, and IoT devices, which can suffer from dropping out during the course of training. For the latter, a notable example is the increasingly popular low Earth orbit (LEO) satellite system that aims at creating a moving infrastructure for seamless global coverage and unlocking real societal benefits by carring out improved Earth observation. While FL is poised to enable more efficient data transmission, the link between individual satellites and ground stations (GS) breaks from time to time due to the visiting pattern of the satellites to the location at which the GS is positioned \cite{razmi2022scheduling}. 

Additionally, HFL and VFL can have greatly different training procedures and client dependencies due to different data structures and, therefore, require different levels of robustness. Each client in HFL trains a complete copy of the global model on their local dataset and sends updated model parameters to a centralized server for aggregation. The effect of having a single client drop-out on the overall training is usually negligible. On the other hand, under VFL, each client holds a specific subset of features and is responsible for training a specific sub-module of the global model, and \emph{the absence of a single client can potentially cripple the entire training process}. Existing VFL research has mainly focused on small client cohorts with reliable communication between clients and the server, and has largely overlooked the realistic but significant issue of client disconnection or drop-out. While the standard way of tackling this problem is to completely discard the training round, which suffers client drop-out, after the pre-defined time-out limit, completely discarding a round can result in a waste of computing power and slower training convergence. Furthermore, the setup of VFL entails the intrinsic risk of privacy leakage as intermediate activations or gradients must be shared between clients during the training process, a process that potentially allows the reconstruction of original raw features ~\cite{dlg, idlg, jin2021cafe,yin2021see}. Therefore, it is critical that any optimization of the VFL training process cannot compromise the security of the system. As such, we introduce \emph{vFedSec}, the novel embedding-padding method, which is a dropout-tolerant method and can largely reduce the waiting time, provide a similar convergence rate as the no drop-out baseline, and maintain the model performance. In this work:
\vspace{-3mm}
\begin{itemize}
    \item We present \emph{vFedSec} - the \textit{first dropout-tolerant} VFL framework that implements an efficient and privacy-preserving training via an architecture-agnostic \emph{Secure Layer}. Section \ref{sec:methodology} describes our proposed design and its core component.
    \vspace{-3mm}
    \item We provide a robust theoretical justification proving that our proposed method does not adversely affect training performance and private information is protected through the \emph{Secure Layer}. 
    \vspace{-3mm}
    \item To validate our claims empirically, we have conducted comprehensive experiments over four varied datasets of differing data volumes and model architectures, as detailed in Section \ref{sec:exp}. The results substantiate that our \emph{vFedSec} imposes negligible overhead in all cases. Moreover, our method demonstrates a remarkable $\geq 690$x speedup and $\geq 9.6$x decrease in communication costs when compared to the resource-intensive homomorphic encryption (HE) techniques, thus affirming its superior computational and communication efficiency. We also demonstrate that with our dropout-tolerance method, we can maintain the training speed while maintaining the model performance.
\end{itemize}

%% file: sections/2_related.tex
\section{Background and Related Work}\label{sec:background}

In the context of an efficient privacy-preserving framework for VFL, several related papers have contributed to the understanding and improvement of privacy, utility, and security aspects in VFL. A comprehensive literature review \cite{vfl} provides an overview of VFL and discusses various protection algorithms. However, it does not specifically address the efficiency aspect of it. Another study \cite{dpvfl} focuses on integrating differential privacy (DP) techniques into VFL, but it does not explicitly cover efficiency either. 
\citet{tradeoffvfl} propose a framework to evaluate the privacy-utility trade-off in VFL, highlighting the limitations of existing approaches like homomorphic encryption (HE) and multi-party computation (MPC) in terms of computation and communication overheads, and \citet{cai2023secure} apply HE to split neural networks (Split NN) for improved security. \citet{zheng2022making} presents a heuristic approach to make split learning resilient to label leakage, albeit at the cost of accuracy. \citet{sun2022label} explore label leakages from forward embeddings and their associated protection methods, which can cause a decrease in accuracy. 
The method proposed by \citet{liu2020asymmetrical} only tries to protect the sample IDs, rather than all the raw private data; \citet{chen2020vafl} perturb local embedding to ensure data privacy and improve communication efficiency, which has strict requirements for the embedding and can impact the overall performance. There are also BlindFL \cite{fu2022blindfl}, ACML \cite{zhang2020acml}, and PrADA \cite{ kang2022prada} which are all homomorphic encryption (HE) based solutions. These approaches often incur significant communication and computation overheads. Moreover, existing approache~\citep{Hardy2017PrivateFL, yang2019quasi} and are restricted to (approximated) linear models and just two parties. 

In the realm of VFL, several studies have delved into the issue of robustness and stragglers problem. \citet{gu2021privacy} present AFSGD-VP, tailored for environments without a central server, ensuring embedded privacy through a tree-structured aggregation. \citet{shi2022} introduce AMVFL, which focuses on asynchronous gradient aggregation for linear and logistic regression, with local embeddings shielded by secret-shared masks. VAFL \cite{chen2020vafl} is designed for intermittently connected clients, incorporating differential privacy (DP) but with some performance compromises compared to our \emph{vFedSec}. FedVS \cite{li2023fedvs} utilizes polynomial interpolation to efficiently handle stragglers, yet it falls short in addressing the drop-out challenge.

%% file: sections/3_method.tex
\section{Methodology} \label{sec:methodology}

\subsection{Problem Settings} \label{sec:setup}

\paragraph{Generalized VFL Framework} 
In our generalized VFL setting, data is partitioned across both feature and sample spaces. Let the complete feature set be denoted as $\mathcal{X}$ and the entire sample set as $\mathcal{D}$. Clients are organized into groups based on the specific features they possess. For group $i$, the dataset $\mathcal{D}_i$ encompasses all samples corresponding to a distinct feature subset $\mathcal{X}_i$, such that $\mathcal{X}_i \cap \mathcal{X}_j = \emptyset$ for all $j \ne i$, and $\mathcal{X} = \bigcup_{i} \mathcal{X}_i$, $\mathcal{D} = \bigcup_{i} \mathcal{D}_i$. Within each group $i$, client $C_i^{(k)}$, the $k$-th client in group $i$, holds a partial dataset $\mathcal{D}_i^{(k)}$ associated with features $\mathcal{X}_i$. These partial datasets contain distinct subsets of samples, ensuring that $\bigcup_{k} \mathcal{D}_i^{(k)} = \mathcal{D}_i$. The first group is considered the active party, comprising a single client, denoted as $C_0$, who possesses the entire label set $\mathcal{Y}$. Other clients are considered to be the passive parties. Importantly, our approach is also compatible with conventional VFL settings (the setting considered in all previous papers), simply by assuming that each group contains only one client. For the sake of simplicity, the active party client $C_0$ is distinguished from other groups, i.e., $C_0$ does not belong to any group. The number of groups is $M$.

\vspace{-2mm}
\paragraph{Classification Task}
In this paper, we focus on a $N$-class classification problem, defined within a compact feature space $\mathcal{X}$ and a discrete label space $\mathcal{Y} = [N]$, where $[N] = \{ 1, \ldots, N \}$ denotes the set of possible class labels. The objective is to train a neural network, represented by a function $f$, parameterized by the model weights $\mathbf{w}$. The performance of this model is evaluated using the loss function $\mathcal{L}(\mathbf{w})$, with the widely adopted cross-entropy loss. Notably, our algorithm exhibits a versatile framework that can be generalized to a variety of other tasks and loss functions, particularly those amenable to split learning. This adaptability allows for the application of our model to a broader range of learning scenarios.

\vspace{-2mm}
\paragraph{Threat Model} 
Our threat model operates under a semi-honest assumption. In this context, the server, denoted as $S$, is trusted not to collude with any clients. However, it may attempt to infer private information from the aggregated data it receives. Likewise, the clients are assumed to follow the protocol honestly, but they may attempt to infer other clients' privacy from all their available information. Our model considers the possibility of collusion among clients. Specifically, one or more clients within the same group or from different groups, can collaborate, posing an increased risk of privacy leakage about other clients' local datasets. This model reflects a realistic scenario in FL environments where the integrity of the server is maintained, but client-to-client trust boundaries are less stringent.

\subsection{Training and Testing Pipeline}
\begin{figure*}[!t]
\centering
\includegraphics[width=0.71\textwidth]{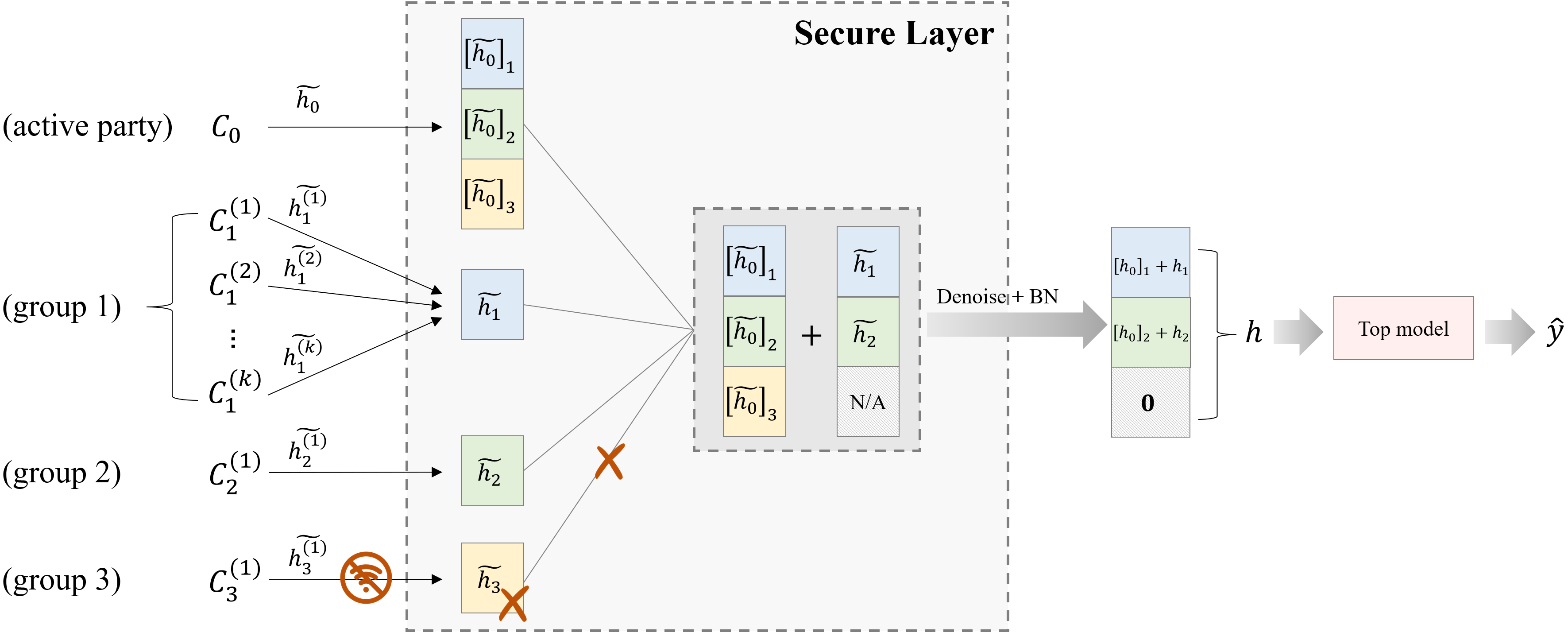}
\caption{Detailed illustration of \emph{vFedSec} with \emph{Secure Layer} with embedding padding techniques during forward pass when drop-out occurs. A client group contains clients with the same set of features, which can be a single client or multiple clients. Each client group is responsible to updating part of the embedding vector, which will be fed to the top model on the server side.  }
\label{fig:forwardpass}
\end{figure*}

This section describe the training and testing pipeline under \emph{vFedSec} protocol. An illustration of it can be found in Figure \ref{fig:protocol} in Appendix \ref{app:protocol}. 

Our framework employs split learning and assumes the establishment of an end-to-end (E2E) encrypted channel between any two clients, enabled by a symmetric encryption algorithm, such as Secure Sockets Layer (SSL). 

The mini-batch selection is assumed to be completed in a privacy-preserving manner, e.g., through a Private Set Intersection (PSI). The specific approach varies based on data sensitivity and the specific scenario. In our experiments, $C_0$ initiates the mini-batch selection, communicating the IDs of chosen samples to the owning clients via the E2E encrypted channel. This ensures that no party gains additional knowledge about samples not in their possession, including sample IDs. In the training phase, labels of the selected samples are sent to the server, but due to the encryption of sample IDs, the server cannot correlate these labels with any specific samples.

\subsubsection{Forward Pass}

During the forward pass, as illustrated in Figure \ref{fig:forwardpass}, $C_0$ inputs raw features $x_0 \in D_0$ into its bottom model $f_0$, parametrized by $\theta_{f_0}$, obtaining the quantized intermediate embedding ${h_0}' = q\left(h_0\right) \in \mathbb{Z}^{B \times H}$, where $h_0 = f_0(x_0)$, $q: \mathbb{R} \rightarrow \mathbb{Z}$ is the quantization function, $B$ is the batch size, and $H$ is the dimension of $h_0$. The exact details regarding quantization can be referred to Section \ref{sec:quant}.

For each client $ C_i^{(k)}$ in group $i$ ($k$ represents the $k$-th client in group $i$), $i=1, 2, \ldots, M$, the process involves passing raw features $x_i^{(k)} \in D_i^{(k)}$ through its bottom model $f_i$, parametrized by $\theta_{f_i}$.  This yields $\Bar{h_i^{(k)}} = f_i\left(x_i^{(k)}\right) \in \mathbb{R}^{Bi^{(k)} \times H_i}$, which is then expanded to $h_i^{(k)} \in \mathbb{R}^{B \times H_i}$ by inserting zeros for missing samples and quantized to ${h_i^{(k)}}' = q\left(h_i^{(k)}\right)$, where $\sum_{i>0}{H_i} = H$. All clients in group $i$ utilize the same bottom model $f_i$ due to shared features space $\mathcal{X}_i$.

The mask embedding of group $i$ for all the selected samples is calculated as ${h_i}' = \sum_k{{h_i^{(k)}}'}$.
For example, consider a scenario where samples $x_1, x_2, x_3$ are selected and group $i$ comprises two clients. Let client $1$ have $x_1$ and client $2$ have $x_2, x_3$. Denote $s_j = q\left(f_i(x_j)\right), j = 1, 2, 3$, as the quantized intermediate embeddings for each sample. The quantized embedding ${h_i}'$ of group $i$ is then derived by adding ${h_i^{(1)}}'$ from client $1$ and ${h_i^{(2)}}'$ from client $2$ as shown below:
\vspace{-2mm}
\begin{equation}
    {h_i^{(1)}}' + 
    {h_i^{(2)}}'
    =
    \begin{pmatrix}
        s_1 \\
        0 \\
        0
    \end{pmatrix}
    +
    \begin{pmatrix}
        0 \\
        s_2 \\
        s_3
    \end{pmatrix}
    =
    \begin{pmatrix}
       s_1 \\
       s_2 \\
       s_3
    \end{pmatrix}
    =
    {h_i}'
\end{equation}

After quantization, $C_0$ generates noise $n_i \in \mathbb{Z}^{B \times H_i}, i=1,2,\ldots, M$, forming the masked intermediate embedding $\Tilde{h_0} = {h_0}' + (n_1^\top, n_2^\top, \ldots, n_{M}^\top)^\top$. Each client $C_i^{(k)}$ in group $i$ similarly produces noise $n_i^{(k)} \in \mathbb{Z}^{B \times H_i}$, resulting in $\Tilde{h_i^{(k)}} = {h_i^{(k)}}' + n_i^{(k)}$. The design ensures that, for all $i$, the aggregate noise $n_i + \sum_{k \in C_i}{n_i^{(k)}}$ is reconstructable on the server side, without revealing individual noises $n_i^{(k)}$ or $n_i$ (as discussed in detail in Section \ref{sec:noise_gen}).

All masked embeddings, including $\Tilde{h_0}$ and $\Tilde{h_i^{(k)}}$ are sent to the server along with the necessary information for noise reconstruction. Let $\Tilde{h_i} = \sum_{k}{\Tilde{h_i^{(k)}}}$ for all $i$. The server reconstructs the aggregated noise $n$ and then calculates the aggregated quantized intermediate embedding $h'$ as follows:
\begin{equation}
    \begin{aligned}
        \small h' &= \Tilde{h} - n = \Tilde{h_0} + \left(\Tilde{h_1}^\top, \ldots, \Tilde{h_M}^\top\right)^\top - n \\
        &= {h_0}' + ({h_1}'^\top, \ldots, {h_M}'^\top)^\top
    \end{aligned}
\end{equation}

where
\begin{equation}
    \small n = \left(\left(n_1 + \sum_{k}{n_1^{(k)}}\right)^\top, \ldots, \left(n_M + \sum_{k}{n_M^{(k)}}\right)^\top\right)^\top 
\end{equation}

The server then dequantizes and feeds the aggregated embedding $h = q^{-1}\left(h'\right)$ into the top model $g$, parametrized by $\theta_g$, to obtain the final output $\hat{y} = g\left(h\right)$.

\subsubsection{Backward Pass}

After computing the output $\hat{y}$ of the top model $g$, the server calculates the loss $\mathcal{L}(\hat{y}, y)$ and its gradient $\nabla \mathcal{L}(\hat{y}, y)$. The server then back-propagates this gradient through the first layer of the top model, updates the model parameters, and sends the intermediate gradients back to the clients.

Upon receiving these gradients, each client proceeds to complete the backpropagation process. The active party, client $C_0$, directly updates its bottom model $f_0$. For a client $C_i^{(k)}$ in group $i$, the update strategy depends on the number of clients in the group. If group $i$ comprises a single client, then $C_i^{(k)}$ updates its bottom model $f_i$ directly. In group $i$ with multiple clients, client $C_i^{(k)}$ computes the model parameters update as $\Delta \theta_i^{(k)} = - \eta \nabla_{\theta_{f_i}} \mathcal{L}(\hat{y}, y)$ and go through the process of quantization to obtain ${\Delta \theta_i^{(k)}}' = q \left( \Delta \theta_i^{(k)}\right)$ like in the forward pass, where $\eta$ denotes the learning rate.

For clients $C_i^{(k)}$ in multi-client group $i$, model updates are masked as $\Tilde{ \Delta \theta_i^{(k)}} = {\Delta \theta_i^{(k)}}' + \epsilon_i^{(k)}$, where $\epsilon_i^{(k)}$ represents noise generated following the same protocol in the forward pass. This ensures that the aggregated noise from all clients' updates in the group $i$, $\sum_{k}{\epsilon_i^{(k)}}$, is reconstructable server-side (See Section \ref{sec:noise_gen}). Each client then sends the masked model update $\Tilde{\Delta \theta_i^{(k)}}$ along with the necessary information for noise reconstruction to the server.

Upon receiving $\Tilde{\Delta \theta_i^{(k)}}$ from all clients in group $i$, the server reconstructs the aggregate noise and obtains the quantized model update as:
\vspace{-3mm}
\begin{equation}
    {\Delta \theta_i}' = \sum_k{\Tilde{\Delta \theta_i^{(k)}}} - \sum_{k}{\epsilon_i^{(k)}} = \sum_k{{\Delta \theta_i^{(k)}}'}
\end{equation}

The server then dequantizes this model update to $\Delta \theta_i = q^{-1}\left( {\Delta \theta_i}'\right)$ and applies it to the bottom model $f_i$. The updated model parameters are subsequently sent to the corresponding clients in the next forward pass.

\subsection{Drop-out Tolenrance} \label{sec:dropout}
Our approach is robust to client drop-out through embedding padding, as illustrated in Figure \ref{fig:forwardpass}. If a client drop-out within group $i$, the server will discard all masked embeddings $\Tilde{h_i^{(k)}}$ from group $i$ and the corresponding segment $[\Tilde{h_0}]_i \in \mathbb{Z}^{B \times H_i}$ of $\Tilde{h_0} \in \mathbb{Z}^{B \times H}$. Consequently, in the aggregated masked embedding $\Tilde{h}$, the segment $[\Tilde{h_0}]_i + \Tilde{h_i}$ is replaced with \texttt{N/A}. 

After noise removal and dequantization of the embedding $h$, the server passes $h$ through a BatchNorm layer. Then, \texttt{N/A} fields within $\textsc{BatchNorm}(h)$ are replaced with zeros. This padded embedding is then input to the top model to generate the final output. This approach ensures the model's resilience to potential disruptions caused by client drop-outs, hence increasing the convergence rate.

\subsection{Noise Generation} \label{sec:noise_gen}

In a system of $N$ clients, for any two clients $u$ and $v$, a pair of noise $n_{u, v}$ and $n_{v, u}$ are generated, satisfying the condition $n_{u, v} = -n_{v, u}$. To facilitate this, a shared seed is established between any two clients, which must be confidential and undisclosed to any third parties. Assuming $u > v$, both client $u$ and $v$ utilize this shared seed to generate identical pseudo-random noise $p$ within the finite field $Z_R^m = \{0, 1, \ldots, R-1\}$, thereby deriving $n_{u, v} = p$ and $n_{v, u} = -p$. The shared seed generation can leverage a list of pre-shared keys, the Elliptic Curve Diffie-Hellman (ECDH) key agreement ~\cite{secagg, ecdh}, or similar methods. For each client $u$, noise $n_{u, v}$ is generated in relation to client $v$ for all $v \ne u$, computing $n_u = \sum_{v \ne u}{n_{u, v}}$. Adding $n_u$ to the client’s intermediate embedding effectively makes it indistinguishable from a random sequence, thereby protecting the client's privacy (see Section \ref{sec:diss1}). Given that $\forall u\ne v, n_{u, v} + n_{v, u} = 0$, the reconstruction and cancellation of the aggregated noises can easily be performed by summation, because the sum of all client noises $n_u$ is zero, as shown below:

\vspace{-6mm}
\begin{equation}
    \sum_u{n_u} = \sum_u{\sum_{v \ne u}{n_{u, v}}} = \frac{1}{2}\sum_{u, v, u \ne v}{\left(n_{u, v} + n_{v, u} \right)} = 0
\end{equation}

Although not used in this paper, the reconstructable sum of noises can alternatively be implemented using techniques such as Lagrange Coded Computing (LCC) \cite{so2022lightsecagg}.

\subsection{Quantization} \label{sec:quant}
Cryptographic operations depend on integer-based computations, thus requiring the quantization of intermediate embeddings. Our method is compatible with any quantization algorithm that facilitates the dequantization of summed quantized vectors. We use a naive quantization algorithm in this work. The quantization process for a real number $x \in \mathbb{R}$ is as follows: define the clipping range as $[-t, t]$ and the target range as $[0, R)$, where $t$ is the clipping threshold and $R$ the size of the target range. The quantized value $x_{\text{quant}} = q(x) = \textsc{round}\left(\frac{\textsc{clip}(x, -t, t) + t}{2t} \times R\right)$ involves clipping $x$ to $[-t, t]$ and applying stochastic rounding. The dequantization is given by $x_{\text{dequant}} = q^{-1}\left(x_{\text{quant}}\right) = \frac{2t}{R}x_{\text{quant}} - t$. Our implementation employs \texttt{float32} to \texttt{uint32} conversion with $t=4$ and $N = 2^{27}$, ensuring minimal impact on model performance due to high precision.

\section{Security Analysis}\label{sec:diss}
\subsection{Privacy Guarantee}\label{sec:diss1}

The thread model we considered is described in Section \ref{sec:setup}. In our security arguments, we show theoretically that the modulus of the intermediate embedding with added noise in integers modulo $R$ ($\mathbb{Z}^m_R$) will be the same as random noise in $\mathbb{Z}^m_R$ for each group of clients.

\begin{lemma}
Let $U_i$ be the set of clients in group $i$ and the active client $C_0$. $\{h'_u\}_{u\in U}$ where $\forall u \in U, h_u \in \mathbb{Z}^m_R$ be the quantized intermediate embedding of each client; $n_{u,v}$ be the random noise associated with client $u$ and $v$ that is uniformly sampled from the finite field $\mathbb{Z}^m_R = \{0, 1, \dots, R -1\}^m$, and $n_{u,v} = - n_{v,u}$. Then we have: intermediate embedding associated with random noises associated with each other clients added: $ h'_u + \sum_{v\in U \setminus \{u\}} n_{u,v} $ has the same distribution as the sum of random noise: $\sum_{u \in U} w_u $ such that $w_u$ is uniformly sampled from $\mathbb{Z}^m_R$ ($w_u \sim \mathbb{Z}^m_R$), provided the sum is the same ($\sum_{u \in U} w_u  =   \sum_{u \in U} h_u (\mathrm{mod} \ R)$).

\begin{align*}
    h_u + \sum_{v\in U \setminus \{u\}} n_{u,v}  \equiv \sum_{u \in U} w_u
\end{align*}



    where `$\equiv$' denotes that the distributions are identical.
\end{lemma}

Proof for the lemma can be found in Appendix \ref{appendix:proof}. The lemma proves that within each set of clients $U$, if clients' values have uniformly random noise added to them, then the resulting values look uniformly random, conditioned on their sum being equal to the sum of the client's intermediate values. In other words, the noises, as explained in Section \ref{sec:noise_gen}, hide all private information about the client's data, apart from the sum. Privacy protection through \emph{Secure Layer} is achieved through the use of masking on the intermediate outputs and gradients before they are sent back to the server, thus preventing the server from using the received information to gain knowledge about the sensitive client information. Therefore, our method protects the data from both data reconstruction and membership inference attacks.

In addition, since \emph{vFedSec} operates for each group of clients as described in Section \ref{sec:dropout}, and the Lemma theoretically proves conditioned on the group of clients, making our method a dropout-tolerant privacy-preserving VFL protocol.

Furthermore, while our current implementation could be vulnerable to active adversaries, our solution can be extrapolated very easily to include \emph{malicious} settings by introducing a public-key infrastructure (PKI) that can verify the identity of the sender \cite{bonawitz2017secagg}. It can thus be further protected from malicious attacks. 

Although our method does not allow exposing secret keys to other parties and demonstrates robustness against collusion between the aggregator and passive parties, in practical settings, the risk of secret key leakage persists, e.g., accidental inappropriate practices of engineers. Thus, to ensure privacy, it is necessary to routinely regenerate keys for symmetric encryption and secure aggregation, specifically, by executing the setup phase after every K iterations, in both the training and testing stages. The value of K can vary in real-world scenarios, but the larger value will inevitably incur higher risks of keys being compromised. In the event of key leakage, an attacker will only have access to a limited amount of information instead of all encrypted information if keys are regenerated periodically.


\subsection{Other considerations}
\textbf{Generalizability:} \emph{vFedSec} provides a generalized protocol for any kind of VFL setups as explained in Section \ref{sec:setup}. We design our protocol \emph{vFedSec} to be able to cope with any kind of vertical FL setups and any kind of model architectures residing both in the local module on the client side and the global module on the server side. Previous papers usually only consider the vertical setup with two parties: one active and one passive \cite{vfl}, or the feature in passive parties cannot overlap \cite{jiang2022vfps, Fang_2021, wei2022vertical}. However, in our general setup, there could be multiple passive parties, and the feature space in passive parties can be overlapped. Also, our protocol can be further expanded to include multiple active parties if needed. In this case, we just need to adapt the mini-batch selection process and the label-sharing process with the server. 

\textbf{Scalability and Efficiency:} Our core solution is agnostic to the number of participating clients and the data partition schemes under the VFL setting. As a result, our solution's scalability is only dependent on the underlying FL framework and on how key generation and key exchange between clients are handled. Also, our solution is scalable and efficient in the sense that, unlike many previous methods explained in Section \ref{sec:background} that utilized HE, we employ lightweight masks through random noise, and the mask can be efficiently reconstructed as explained in Section \ref{sec:noise_gen}. 

%% file: sections/4_exp.tex
\section{Experiments}\label{sec:exp}

\subsection{Experimental Protocol}\label{sec:expsetup}

\textbf{Datasets:} Experiments are conducted over four datasets: Bank Marketing dataset \cite{bankdataset}, Adult Income dataset \cite{adultincome}, EMNIST \cite{cohen2017emnist}, and Fashion-MNIST \cite{xiao2017fashion}. 

\textbf{Drop-out Simulation.} To simulate the drop-out in the experiments, we introduce two different coefficients. We simulate the drop-out as a random process in our experiments. The first one refers to the drop-out probability of a given round. If the probability is $0.3$, it means that there is a 30$\%$ chance that at least one client will drop out in this particular round. Another coefficient refers to the proportion of clients in the client pool dropping out. If the proportion is $0.1$, it means that there are random $10\%$ of clients who will drop out in this particular round. We also consider the baseline case, which we refer to as the `discard' case, which means that if there exists a client dropping out, we discard this round completely. For example, if we say discard ($0.3$,$0.1$) it refers to the discard case with $0.3$ drop-out probability and $0.1$ proportion of clients dropping out. We aim to simulate the situation when the drop-out probability is relatively high to demonstrate the superiority of our method compared with simply discarding the current communication round if the drop-out happens. In our experiments, we choose two different values for the drop-out probability ($0.3$ and $0.4$) and choose $0.1$ as the client drop-out proportion.

Other details regarding the datasets, specific partition, model architecture, and the details of overhead profiling can be found in Appendix \ref{app:expsetup}. 



\input{sections/7_huge_table}

\subsection{Efficientcy Results}
\label{sec:overhead}
\emph{vFedSec} is agnostic to the number of participating clients, and is efficient in terms of both computation and communication. In this section, we conduct experiments over four datasets to measure both the computation and the communication cost of \emph{vFedSec} training. 

The computation cost is measured through CPU time, and the communication cost is measured through the transmission size. We also measure the overhead cost that shows the extra CPU time or communication compared to unsecured VFL training.  Communication overheads are from the length of ciphertext (encrypted Sample IDs) over plaintext and the exchange of public keys. Computation overheads are caused by creating key pairs and shared secrets, noise generation and reconstruction, and encryption and decryption. As transferring parameters/gradients and updating models are necessary for VFL training and split learning, they are not counted as overheads. All experiments are reported with 1 setup phase and 5 training rounds, and each experiment is repeated 10 times, and averages are reported.

As mentioned in Section \ref{sec:diss1}, in practice, each party should create new key pairs routinely to mitigate the risk of adversaries from accessing confidential information in the event of secret key leakage. In our experiments, the key pairs and the shared secrets will be re-generated for every 5 iterations. 

Table \ref{tab:cpu_time} reports the CPU time as a measure to show the computation cost using \emph{vFedSec}. The equivalence of certain total times to overhead times can be attributed to the substantial overhead of HE, which overshadows other computational processes. The CPU time is reported separately for the active party and passive parties. The overhead columns show extra CPU time compared with unsecured VFL training. We use HE functions from Python module \emph{Phe} \cite{phe} and \emph{SEAL-Python} \cite{sealpy}. \emph{Phe} module implements the Pallier cryptosystem in Python, and \emph{SEAL-Python} creates Python bindings for APIs in Microsoft SEAL \cite{sealcrypto} using Pybind11 \cite{pybind11}. Considering that the protocols proposed by many HE VFL papers, such as BlindFL\cite{fu2022blindfl}, ACML\cite{zhang2020acml}, and PrADA\cite{kang2022prada}, they entail encrypting a matrix and computing matrix multiplication between an encrypted matrix and a plaintext matrix, we estimated the overhead of above operations through (1) encrypt a weight matrix in each aggregation; (2) compute a matmul between a plaintext matrix and an encrypted weight matrix in each aggregation. It is worth noting that, in reality, most HE VFL protocols require multiple rounds of communication, repetitively encrypt/decrypt matrices, and conduct matmul. But, in our estimated cases, we only encrypt and matmul once in each forward pass, and exclude the overhead of decryption (as decryption may happen on the server side). Thus, our estimation is the minimum overhead incurred by HE VFL protocols, while the real overheads of adopting those protocols can be multiple times larger than our estimation if using the same HE libraries. 

Table \ref{tab:communication} shows the transmission size and is also demonstrated on both the active and passive parties. As demonstrated in both tables, the overhead accounts for a relatively small part of the total amount, in both CPU time and communication size. The CPU time overhead is caused by parties adding masks to their original output and encryption/decryption of sample IDs. The communication overhead is introduced by broadcasting encrypted sample IDs, which are larger than plain text. As the masks can be reconstructed by summation, the unmasking process is very efficient.

\input{sections/4_2_dropoutplot}

Also, as reported in Table \ref{tab:complexity}, the client-side overheads of \emph{vFedSec} grow linearly with increasing the number of clients and the batch size. The server-side computation and storage overheads are negligible, and the communication overheads are incurred by broadcasting encrypted sample IDs to all clients, instead of only forwarding sample IDs to the sample holders. More results can be found in Appendix \ref{app:res}.

\subsection{Drop-out Results}
As mentioned in the introduction, clients in real-world scenarios could drop out from the current round of training. In contrast to the standard solution (i.e. completely discard this round of training after the pre-defined time-out limit), which yields slow convergence due to waste of computing power, \emph{vFedSec} introduces the novel dropout-tolerant embedding padding method, which greatly reduces the waiting time when drop-out occurs and eliminates the gap between the drop-out scenario and no drop-out baseline in terms of convergence rate while maintaining the model performance. 

Figure \ref{fig:dropout} shows the results for Fashion-MNIST and EMNIST under different drop-out scenarios. The plots show that our method can maintain the model performance with a similar convergence rate under different feature partition setups, but the discard baseline fails to do so. Similar results can be found in Table \ref{tab:dropout} for other datasets too in Appendix \ref{app:dropout}. These results also show that with more partitions, the discard baselines perform worse while our method still holds the same level of performance. This is mainly because, when drop-out occurs, the part that requires padding reduces as the number of partitions increases. Hence, it might have less impact on the model performance.
\subsection{Ablation Study}

\begin{table}[!h]
\centering
    \scalebox{0.8}{
 \begin{tabular}{lcc}
        \toprule
        & \textbf{Client} & \textbf{Server} \\ 
        \midrule
        \textbf{Computation} & $O(hN+B)$ & $O(1)$ \\
        \textbf{Communication} & $O(N + B)$ & $O(BN)$ \\
        \textbf{Storage} & $O(N)$ & $O(1)$ \\
        \bottomrule
    \end{tabular}
    }
    \caption
      { \small
        Overhead summary of \emph{vFedSec} in one training/testing round. $h$ is the size of intermediate output, $B$ is the batch size, and $N$ is the number of clients. 
      }
      \label{tab:complexity}%
\end{table}

\begin{figure}[!h]
\centering
    \includegraphics[width=0.7\linewidth]{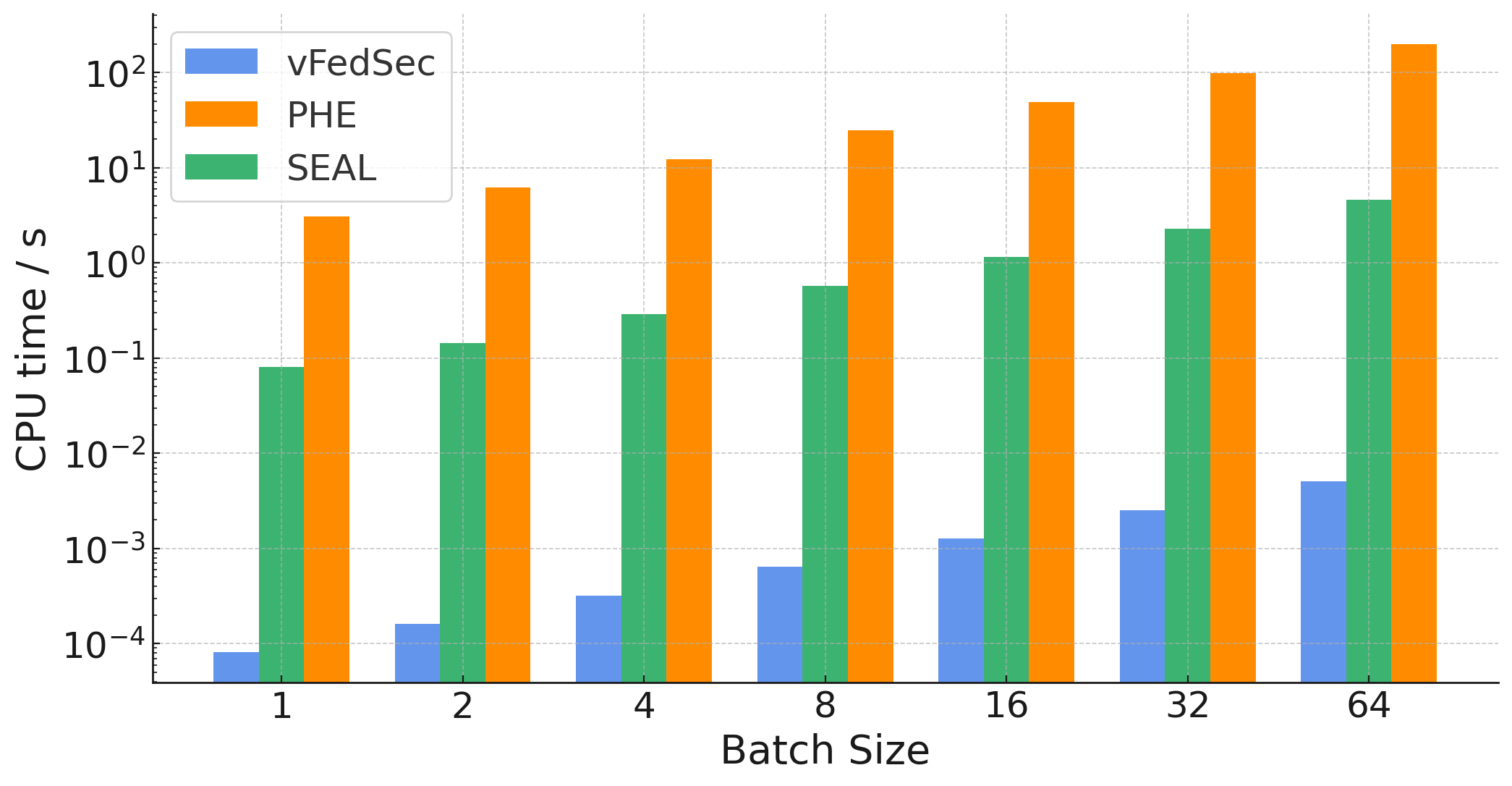}
\caption{\small Comparison of average CPU time for different batch sizes, using \emph{vFedSec} and HE from Phe and SEAL-Python. The Y-axis is in the log scale. The results are from 10 experiments.}
\label{fig:cmp}
\end{figure}

We demonstrate the efficiency of \emph{vFedSec} through an ablation study to compare our method and widely adopted method HE. The experiments compare how \emph{vFedSec} and HE process matrix multiplications. 

Assume the input tensor is of size (Batch size, 8), and the weight tenor is (8, 8). Tensor shapes in the comparison are smaller than tensors used by a passive party in experiments. Given that the HE libraries do not support matrix operations, both \emph{vFedSec} and HE implementations are not optimized by any Python modules, such as \emph{numpy}. The HE implementation in this comparison inevitably involves nested Python loops. For large matrices or more advanced operations, implementations in other languages, e.g., C++, C\#, are more suitable. The results can be found in Fig. \ref{fig:cmp}, which clearly shows the efficiency of our method. It indicates that our approach yields a \num{9.1e2} $\sim$ \num{3.8e4} times speedup when evaluating the computational overheads attributed solely to our method in contrast to HE.

%% file: sections/7_huge_table.tex
\begin{table*}[ht!]
\centering
\caption{Results on the CPU time (in seconds). Note that PHE package is less optimized than SEAL. It is generally not considered in real-world applications.}
\scalebox{0.67}{
\begin{tabular}{c|cccccccc}
\toprule
\multicolumn{1}{c}{} & \multicolumn{4}{c}{Active Party CPU time (s)} & \multicolumn{4}{c}{Passive Party CPU time (s)} \\
\cmidrule(lr){2-5} \cmidrule(lr){6-9}
\multicolumn{1}{c}{} & \multicolumn{2}{c}{Training phase} & \multicolumn{2}{c}{Testing phase} & \multicolumn{2}{c}{Training phase} & \multicolumn{2}{c}{Testing phase} \\
\cmidrule(lr){2-3} \cmidrule(lr){4-5} \cmidrule(lr){6-7} \cmidrule(lr){8-9}
\multicolumn{1}{c|}{Dataset} & Total & Overhead & Total & Overhead & Total & Overhead & Total & Overhead \\
\midrule
\textbf{Bank(Ours)} & \textbf{1.044} & \textbf{0.1881} & \textbf{0.3057} & \textbf{0.1864} & \textbf{0.1633} & \textbf{0.1089} & \textbf{0.1363} & \textbf{0.1075} \\
Bank(SEAL)      & $1.075 \times 10^3$ & $1.075 \times 10^3$ & $1.075 \times 10^3$ & $1.075 \times 10^3$ & $1.113 \times 10^2$ & $1.113 \times 10^2$ & $1.113 \times 10^2$ & $1.113 \times 10^2$ \\
Bank(PHE)       & $1.565 \times 10^4$ & $1.565 \times 10^4$ & $1.565 \times 10^4$ & $1.565 \times 10^4$ & $1.993 \times 10^3$ & $1.993 \times 10^3$ & $1.993 \times 10^3$ & $1.993 \times 10^3$ \\
\midrule
\textbf{Income(Ours)}  & \textbf{0.7379} & \textbf{0.1909} & \textbf{0.2738} & \textbf{0.1889} & \textbf{0.1588} & \textbf{0.1143} & \textbf{0.1413} & \textbf{0.1124} \\
Income(SEAL) & $5.093 \times 10^2$ & $5.092 \times 10^2$ & $5.093 \times 10^2$ & $5.092 \times 10^2$ & $3.885 \times 10^2$ & $3.884 \times 10^2$ & $3.884 \times 10^2$ & $3.884 \times 10^2$ \\
Income(PHE)  & $7.413 \times 10^3$ & $7.413 \times 10^3$ & $7.413 \times 10^3$ & $7.413 \times 10^3$ & $6.853 \times 10^3$ & $6.853 \times 10^3$ & $6.853 \times 10^3$ & $6.853 \times 10^3$ \\
\midrule
\textbf{EMNIST(Ours)} & \textbf{1.040} & \textbf{0.3334} & \textbf{0.6671} & \textbf{0.3300} & \textbf{0.8538} & \textbf{0.2314} & \textbf{0.5386} & \textbf{0.2314} \\
EMNIST(SEAL)   & $7.918 \times 10^4$ & $7.918 \times 10^4$ & $7.918 \times 10^4$ & $7.918 \times 10^4$ & $7.918 \times 10^4$ & $7.918 \times 10^4$ & $7.918 \times 10^4$ & $7.918 \times 10^4$ \\
EMNIST(PHE)    & $1.153 \times 10^6$ & $1.153 \times 10^6$ & $1.153 \times 10^6$ & $1.153 \times 10^6$ & $1.153 \times 10^6$ & $1.153 \times 10^6$ & $1.153 \times 10^6$ & $1.153 \times 10^6$ \\
\midrule
\textbf{Fashion-MNIST(Ours)} & \textbf{1.016} & \textbf{0.3240} & \textbf{0.6286} & \textbf{0.3189} & \textbf{0.8504} & \textbf{0.2301} & \textbf{0.5282} & \textbf{0.2300} \\
Fashion-MNIST(SEAL)   & $7.918 \times 10^4$ & $7.918 \times 10^4$ & $7.918 \times 10^4$ & $7.918 \times 10^4$ & $7.918 \times 10^4$ & $7.918 \times 10^4$ & $7.918 \times 10^4$ & $7.918 \times 10^4$ \\
Fashion-MNIST(PHE)    & $1.153 \times 10^6$ & $1.153 \times 10^6$ & $1.153 \times 10^6$ & $1.153 \times 10^6$ & $1.153 \times 10^6$ & $1.153 \times 10^6$ & $1.153 \times 10^6$ & $1.153 \times 10^6$ \\
\bottomrule
\end{tabular}

}

\captionsetup{font=small,labelfont=bf}
\label{tab:cpu_time}
\end{table*}

\begin{table*}[ht!]
\centering
\caption{Results on the communication both in size (MB).}
\scalebox{0.67}{
\begin{tabular}{c|p{1.69cm}p{1.69cm}p{1.69cm}p{1.69cm}p{1.69cm}p{1.69cm}p{1.69cm}p{1.69cm}}
\toprule
\multicolumn{1}{c}{} & \multicolumn{4}{c}{Active Party Data Transmission (MB)} & \multicolumn{4}{c}{Passive Party Data Transmission (MB)} \\
\cmidrule(lr){2-5} \cmidrule(lr){6-9}
\multicolumn{1}{c}{} & \multicolumn{2}{c}{Training phase} & \multicolumn{2}{c}{Testing phase} & \multicolumn{2}{c}{Training phase} & \multicolumn{2}{c}{Testing phase} \\
\cmidrule(lr){2-3} \cmidrule(lr){4-5} \cmidrule(lr){6-7} \cmidrule(lr){8-9}
\multicolumn{1}{c|}{Dataset} & Total & Overhead & Total & Overhead & Total & Overhead & Total & Overhead \\
\midrule
\textbf{Bank(Ours)}&\textbf{0.88}&\textbf{0.14}&\textbf{0.57}&\textbf{0.14}&\textbf{0.79}&\textbf{0.13}&\textbf{0.44}&\textbf{0.13}\\
Bank(SEAL)&8.41&7.67&8.10&7.67&8.32&7.57&7.97&7.57\\
Bank(PHE)&68.70&67.95&68.38&67.95&68.60&67.85&68.26&67.85\\

\midrule

\textbf{Income(Ours)}&\textbf{0.88}&\textbf{0.14}&\textbf{0.57}&\textbf{0.14}&\textbf{0.85}&\textbf{0.13}&\textbf{0.44}&\textbf{0.13}\\
Income(SEAL)&8.41&7.67&8.10&7.67&8.39&7.57&7.97&7.57\\
Income(PHE)&68.70&67.95&68.38&67.95&68.67&67.85&68.26&67.85\\



\midrule

\textbf{EMNIST(Ours)}&\textbf{6.24}&\textbf{0.21}&\textbf{3.31}&\textbf{0.21}&\textbf{5.99}&\textbf{0.05}&\textbf{3.06}&\textbf{0.05}\\
EMNIST(SEAL)&76.85&70.82&73.92&70.82&76.60&70.66&73.67&70.66\\
EMNIST(PHE)&641.98&635.95&639.05&635.95&641.73&635.79&638.80&635.79\\

\midrule

\textbf{Fashion-MNIST(Ours)}&\textbf{6.24}&\textbf{0.21}&\textbf{3.31}&\textbf{0.21}&\textbf{5.99}&\textbf{0.05}&\textbf{3.06}&\textbf{0.05}\\
Fashion-MNIST(SEAL)&76.85&70.82&73.92&70.82&76.60&70.66&73.67&70.66\\
Fashion-MNIST(PHE)&641.98&635.95&639.05&635.95&641.73&635.79&638.80&635.79\\

\bottomrule
\end{tabular}
}
\captionsetup{font=small,labelfont=bf}
\label{tab:communication}
\end{table*}

%% file: sections/4_2_dropoutplot.tex
\begin{figure*}[!t]
    \centering
    \subfigure[Fashion-MNIST 4 partitions]{\includegraphics[width=0.24\textwidth]{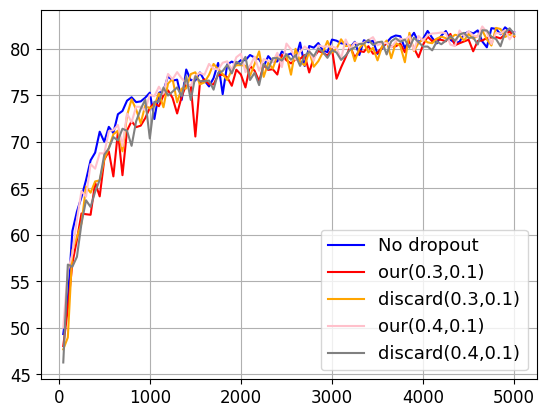}} 
    \subfigure[Fashion-MNIST 8 partitions]{\includegraphics[width=0.24\textwidth]{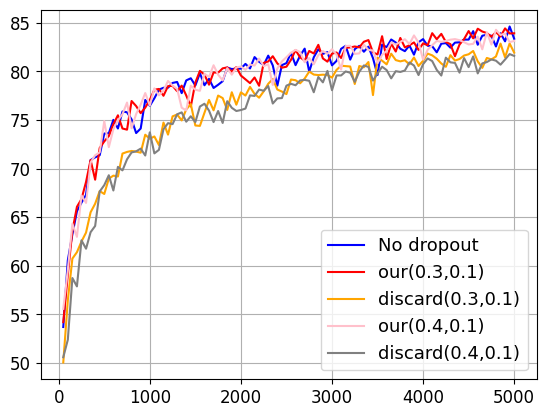}}
    \subfigure[EMNIST 4 partitions]{\includegraphics[width=0.24\textwidth]{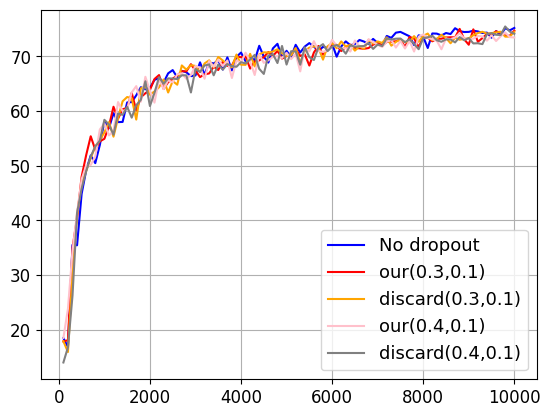}} 
    \subfigure[EMNIST 8 partitions]{\includegraphics[width=0.24\textwidth]{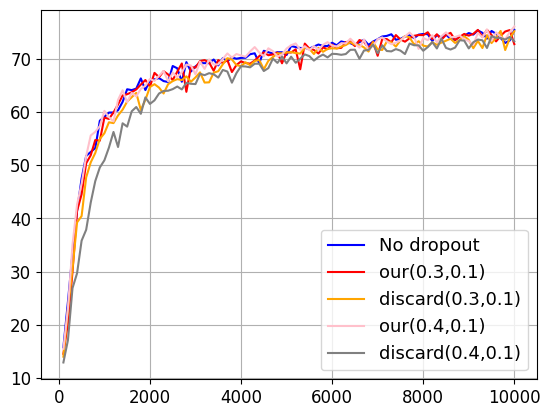}} 

    \caption{Plots to show the performance in different drop-out setups. The experimental protocol can be found in Section \ref{sec:expsetup}. The partition number refers to the number of partitioned feature spaces.}
    \label{fig:dropout}
\end{figure*}

%% file: sections/5_conclusion.tex
\section{Conclusion}
\label{sec:conc}
In this work, we tackle the challenge of privacy-preserving training in vertical FL settings. We propose \emph{vFedSec}, the first framework that works securely, robustly, and efficiently under the most generalized vertical FL setting and unreliable client connectivity. We introduce innovative \emph{Secure Layer} and provide theoretical proof that its design attains enhanced security while maintaining model performance. We further demonstrate through empirical results that our method, while effectively and efficiently protecting sensitive private information, also maintains model performance in the event of unreliable connectivity and a high drop-out rate. Furthermore, this novel method not only provides a significantly faster convergence rate compared with the standard method of tacking client drop-out, but also eliminates the gap between the drop-out scenario and no drop-out baseline in terms of convergence rate while maintaining the model performance. Finally, we also benchmarked our method against HE-based solutions using two HE libraries, which indicated a remarkable reduction in computational costs by at least $690\times$, and a significant decrease in communication costs by more than $9.6\times$.

%% file: sections/99_appendix.tex
\section{Illustration of our protocol} \label{app:protocol}
Figure \ref{fig:protocol} illustrate on a high-level of our protocol.
\input{sections/6_graph}

\section{Experiment Setup} \label{app:expsetup}


\textbf{Banking Datasets:} The Bank dataset is related to the direct marketing campaigns of a Portuguese banking institution. It contains $45,211$ rows and $18$ columns ordered by date. We keep the \texttt{housing}, \texttt{loan}, \texttt{contact}, \texttt{day}, \texttt{month},
\texttt{campaign}, \texttt{pdays}, \texttt{previous}, \texttt{poutcome} features in the \emph{active party}. Features \texttt{default}, \texttt{balance} are seen in Client $1$ and $2$, while \texttt{age}, \texttt{job}, \texttt{marital}, \texttt{education} are kept in Client $3$ and $4$. In this case, Client $1$ and $2$ are in the same group, while Client $3$ and $4$ are in the same group.

\textbf{Adult Income Dataset:} The Adult Income dataset is a classification dataset aiming to predict whether the income exceeds 50K a year based on census data. It contains $48,842$ and $14$ columns. We keep features \texttt{workclass}, \texttt{occupation}, \texttt{capital-gain}, \texttt{capital-loss}, \texttt{hours-per-week} in the active party and \texttt{race}, \texttt{marital-status}, \texttt{relationship}, \texttt{age} \texttt{gender}, \texttt{native-country} are kept by Client $1$ and $2$, while \texttt{education} is held by Client $3$ and $4$. In this case, Client $1$ and $2$ are in the same group, while Client $3$ and $4$ are in the same group.
%


\textbf{EMNIST:} The EMNIST balanced dataset is an expansion of the original MNIST to include handwritten characters, which contains $131,600$ images from 47 balanced classes. The size of images in the dataset is $28\times 28$. To emulate VFL, each image is equally partitioned into 4 "slices", sized $7 \times 28$, and held by 4 different parties, i.e., one active party and three passive parties.

\textbf{Fashion-MNIST:} The Fashion-MNIST dataset, i.e., the Fashion MNIST dataset, comprises $70,000$ grayscale images of clothing items and accessories. The image size of this dataset is the same as that of EMNIST, and hence we adopt the same approach to partitioning the dataset.

\textbf{Drop-out partitions:} To simulate the drop-out performance for different numbers of partitions. We consider randomly partition features for different numbers of partitions for each client. In this case, each group only contains 1 client.

\textbf{Model Architecture.} We consider different model architectures for each dataset. Features and models are partitioned among different parties in experimental settings. For the Banking dataset, the active party used Linear(57, 64); Client $1$ and $2$ used unbiased Linear(3, 64); Client $3$ and $4$ used unbiased Linear(20, 64). The three local modules combined are equivalent to Linear(80, 64). The global module owned by the aggregator comprised Linear(64, 1). For the Adult Income dataset, the active party, Client $1$ and $2$, and Client $3$ and $4$ possessed Linear(27, 64), unbiased Linear(63, 64), and unbiased Linear(16, 64) respectively. The three are equivalent to Linear(106, 64). The global module had Linear(64, 1). We conduct experiments with CNN/MLP models on EMNIST/FMNIST datasets. For CNN on EMNIST, each passive party has Conv3x3 with 32 channels followed by Conv1x1 with 64 channels; each conv layer is followed by BatchNorm, ReLU, and MaxPool2x2; the active party has Linear(600, 120) and Linear(120, 47). For MLP on EMNIST, each passive party has Linear(196, 32) and Linear(32, 128); the active party has Linear(128, 256), Linear(256, 128), Liner(128, 64), and Linear(64, 47). The CNN and MLP used for FMNIST are the same as those for EMNIST; only the number of classes differs. All models in the drop-out simulations are CNN/MLP models for consistency.

\textbf{Overhead profiling.} To quantify the additional communication and computation expenses associated with our method, we measured the total CPU time and the total amount of upload/ download bytes, including their respective overheads, compared to the unsecured split learning for VFL. In practice, parties should create new key pairs routinely, which mitigates the risk of adversaries potentially accessing all confidential information in the event of secret keys used in some iterations being unexpectedly leaked. In experiments, the key pairs and the shared secrets will be regenerated for every 5 iterations.

We used a learning rate of 0.01 and a batch size of 256. We applied ReLU activation to all layers except the output layer.

\section{Proof for the Lemma} \label{appendix:proof}
\begin{proof}

One thing we would like to make clear is that everything related to the cryptography is based on the positive integers. Therefore, before adding masks, we need to perform quantization to the intermediate output or gradients and then make them into the positive integers space, and everything is operated in the space of the integers modulo R ($\mathbb{Z}_R$).

According to distributive law, we have:
\begin{equation}
    (a \mod R + b \mod R) \mod R = (a+b) \mod R
\end{equation}

The number R is chosen that both the sum of intermediate output or gradient and the mask generated by the PRG is in the integer space $Z_R$. 

We can prove by induction on the number of clients (N) in set $U$.

First, assume there are 2 clients in the system: client 1 and client 2. We let the mask generated by the $\mathrm{PRG}(ss_{12})$ be $M$, which is uniformly distributed in the integer space $Z_R$. For a given intermediate output value $v_1$ and $v_2$ for client 1 and 2, the output after the mask is ($v_1 + M$) and ($v_2 - M$), so according to the distributive law, the results ($v_1 + M$) and ($v_2 - M$) are both uniformly distributed in $Z_R$.

Then, by induction and distributive law, with more clients in the system, since we are adding more uniformly distributed masks into the equation, the results are still uniformly distributed in $Z_R$. 
\end{proof}

\section{Additional Results} \label{app:res}

Table \ref{tab:cpu_time_app} and \ref{tab:communication_app} provides additional overhead results. Both tables show that our method provides significantly less overhead both in terms of computation and communication costs. 

\begin{table*}[h!]
\centering
\caption{Results of MLP on the CPU time (in seconds)}
\scalebox{0.68}{
\begin{tabular}{c|cccccccc}
\toprule
\multicolumn{1}{c}{} & \multicolumn{4}{c}{Active Party CPU time (s)} & \multicolumn{4}{c}{Passive Party CPU time (s)} \\
\cmidrule(lr){2-5} \cmidrule(lr){6-9}
\multicolumn{1}{c}{} & \multicolumn{2}{c}{Training phase} & \multicolumn{2}{c}{Testing phase} & \multicolumn{2}{c}{Training phase} & \multicolumn{2}{c}{Testing phase} \\
\cmidrule(lr){2-3} \cmidrule(lr){4-5} \cmidrule(lr){6-7} \cmidrule(lr){8-9}
\multicolumn{1}{c|}{Dataset} & Total & Overhead & Total & Overhead & Total & Overhead & Total & Overhead \\
\midrule
\textbf{EMNIST(Ours)} & \textbf{0.5551} & \textbf{0.3282} & \textbf{0.4786} & \textbf{0.3277} & \textbf{0.4000} & \textbf{0.1949} & \textbf{0.3423} & \textbf{0.1949} \\
EMNIST(SEAL)   & $1.207 \times 10^3$ & $1.207 \times 10^3$ & $1.207 \times 10^3$ & $1.207 \times 10^3$ & $1.207 \times 10^3$ & $1.207 \times 10^3$ & $1.207 \times 10^3$ & $1.207 \times 10^3$ \\
EMNIST(PHE)    & $1.757 \times 10^4$ & $1.757 \times 10^4$ & $1.757 \times 10^4$ & $1.757 \times 10^4$ & $1.757 \times 10^4$ & $1.757 \times 10^4$ & $1.757 \times 10^4$ & $1.757 \times 10^4$ \\
\midrule
\textbf{FMNIST(Ours)} & \textbf{0.5438} & \textbf{0.3304} & \textbf{0.4723} & \textbf{0.3301} & \textbf{0.3947} & \textbf{0.2003} & \textbf{0.3338} & \textbf{0.2003} \\
FMNIST(SEAL)   & $1.207 \times 10^3$ & $1.207 \times 10^3$ & $1.207 \times 10^3$ & $1.207 \times 10^3$ & $1.207 \times 10^3$ & $1.207 \times 10^3$ & $1.207 \times 10^3$ & $1.207 \times 10^3$ \\
FMNIST(PHE)    & $1.757 \times 10^4$ & $1.757 \times 10^4$ & $1.757 \times 10^4$ & $1.757 \times 10^4$ & $1.757 \times 10^4$ & $1.757 \times 10^4$ & $1.757 \times 10^4$ & $1.757 \times 10^4$ \\
\bottomrule
\end{tabular}

}

\captionsetup{font=small,labelfont=bf}
\label{tab:cpu_time_app}
\end{table*}

\begin{table*}[h!]
\centering
\caption{Results of MLP on the communication both in size (MB).}
\scalebox{0.8}{
\begin{tabular}{c|cccccccc}
\toprule
\multicolumn{1}{c}{} & \multicolumn{4}{c}{Active Party Data Transmission (MB)} & \multicolumn{4}{c}{Passive Party Data Transmission (MB)} \\
\cmidrule(lr){2-5} \cmidrule(lr){6-9}
\multicolumn{1}{c}{} & \multicolumn{2}{c}{Training phase} & \multicolumn{2}{c}{Testing phase} & \multicolumn{2}{c}{Training phase} & \multicolumn{2}{c}{Testing phase} \\
\cmidrule(lr){2-3} \cmidrule(lr){4-5} \cmidrule(lr){6-7} \cmidrule(lr){8-9}
\multicolumn{1}{c|}{Dataset} & Total & Overhead & Total & Overhead & Total & Overhead & Total & Overhead \\
\midrule
\textbf{EMNIST(Ours)} & \textbf{1.63} & \textbf{0.21} & \textbf{1.01} & \textbf{0.21} & \textbf{1.38} & \textbf{0.05} & \textbf{0.75} & \textbf{0.05} \\
EMNIST(SEAL)&16.70&15.27&16.07&15.27&16.44&15.12&15.82&15.12\\
EMNIST(PHE)&137.26&135.83&136.63&135.83&137.01&135.68&136.38&135.68\\

\midrule

\textbf{FMNIST(Ours)} & \textbf{1.63} & \textbf{0.21} & \textbf{1.01} & \textbf{0.21} & \textbf{1.38} & \textbf{0.05} & \textbf{0.75} & \textbf{0.05} \\
FMNIST(SEAL)&16.70&15.27&16.07&15.27&16.44&15.12&15.82&15.12\\
FMNIST(PHE)&137.26&135.83&136.63&135.83&137.01&135.68&136.38&135.68\\
\bottomrule
\end{tabular}
}

\captionsetup{font=small,labelfont=bf}
\label{tab:communication_app}
\end{table*}

\section{Addtional Drop-out results} \label{app:dropout}
\input{sections/4_dropouttable}

Table \ref{tab:dropout} shows the performance in different drop-out setups on Bank marketing and Adult income dataset. Since both datasets are binary classification task, we report the AUC rather than the simple accuracy. The experimental protocol can be found in Section \ref{sec:expsetup}. The table demonstrates that our method can endure a high drop-out rate with minimal performance degradation while maintaining the training speed. All experiments are repeated $5$ times, and the average is reported.

%% file: sections/6_graph.tex
\begin{figure}[h]
\centering

\includegraphics[width=0.9\linewidth]{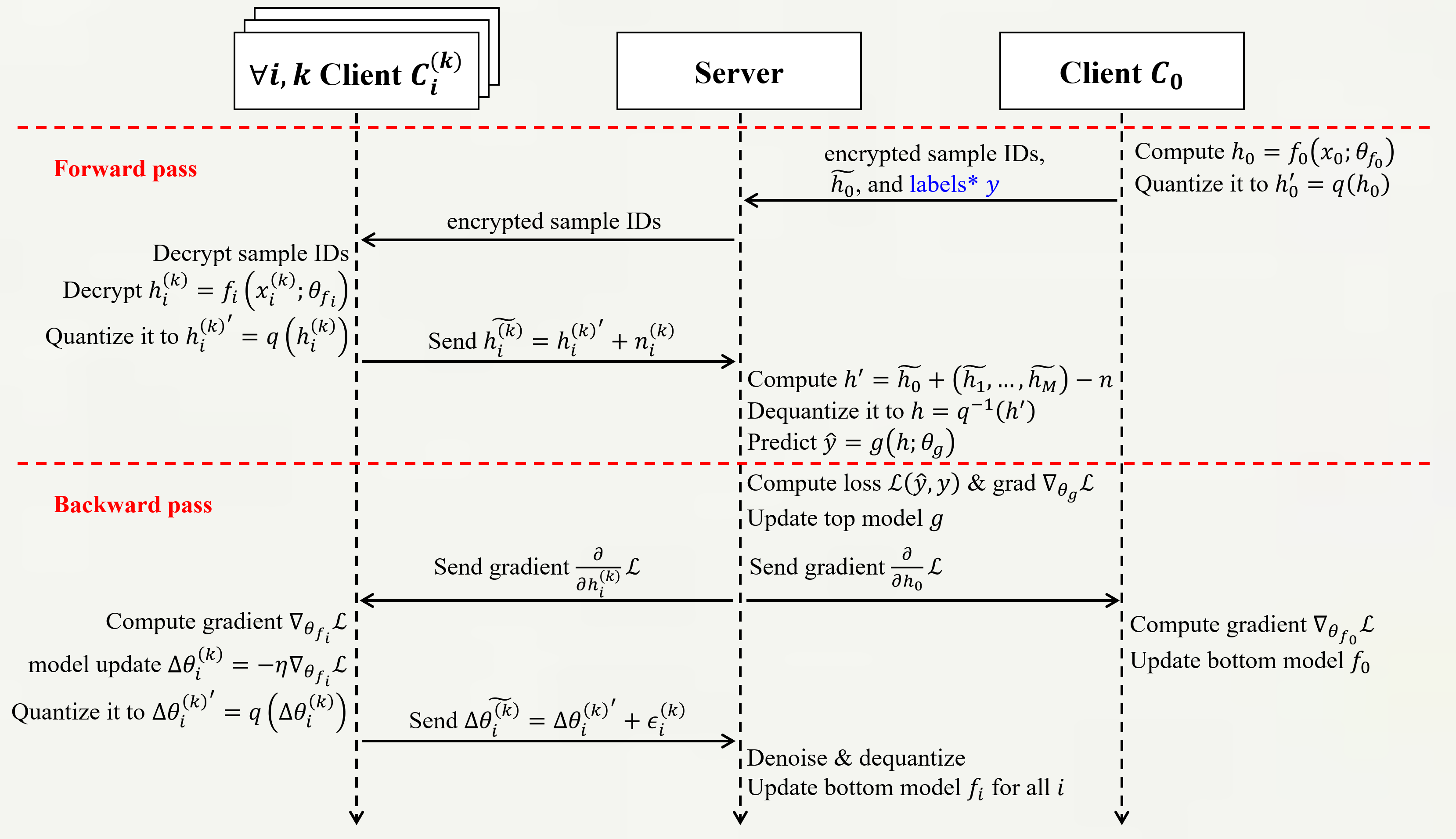} 
\caption{A high-level view of our protocol.  \color{blue} Labels \color{black} in the forward pass section are only required in training. }
\label{fig:protocol}
\end{figure}

%% file: sections/4_dropouttable.tex
\begin{table*}[h]
    \caption{\small Results to show the performance in different drop-out setups on Bank Marketing and Adult Income datasets. The experimental protocol can be found in Section \ref{sec:expsetup}. We report the AUC at the round number 30 and 50, to demonstrate the faster convergence rate when using our method.
    }

    \label{tab:dropout}
    \centering
    \scalebox{0.8}{
    \begin{tabular}{l cccc | cccc}
    \toprule
    Dataset & Partition & Setup & AUC(30) & AUC(50) & Partition & Setup & AUC(30) & AUC(50)\\
    \midrule
    \multirow{4}{*}{Bank}  & \multirow{4}{*}{5}& Discard(0.3,0.1) & 82.19$\%$ & 82.19$\%$ & \multirow{4}{*}{8}&Discard(0.3,0.1) & 74.54$\%$ & 79.15$\%$ \\
    & & Our(0.3,0.1) & 83.25$\%$ & 83.28$\%$ & &Our(0.3,0.1) & 82.49$\%$ & 82.53$\%$ \\
    & & Discard(0.4,0.1) & 62.73$\%$ &  82.20$\%$&& Discard(0.4,0.1) & 73.27$\%$& 81.56$\%$ \\
    & & Our(0.4,0.1) &  80.19$\%$& 82.37$\%$ &&Our(0.4,0.1) & 78.55$\%$ & 82.61$\%$ \\   
    \midrule

    \multirow{4}{*}{Income}  & \multirow{4}{*}{5}& Discard(0.3,0.1) &79.88$\%$ & 81.38$\%$ & \multirow{4}{*}{8} &Discard(0.3,0.1) & 76.15$\%$ & 81.08$\%$ \\
    & & Our(0.3,0.1) & 82.63$\%$ & 83.28$\%$ & &Our(0.3,0.1) & 79.95$\%$ & 82.32$\%$\\
    & & Discard(0.4,0.1) & 74.41$\%$ & 78.63$\%$& &Discard(0.4,0.1) & 75.36$\%$ & 78.88$\%$ \\
    & & Our(0.4,0.1) & 80.18$\%$ & 82.45$\%$& &Our(0.4,0.1) & 80.18$\%$ & 82.45$\%$ \\   
    \bottomrule
    \end{tabular}
}
\end{table*}